\pdfoutput=1
\documentclass[10pt,aps,prb,amsmath,amssymb,twocolumn,%
                letterpaper,footinbib,%
                superscriptaddress,citeautoscript,%
                floatfix]{revtex4-1}

\usepackage[usenames,dvipsnames]{xcolor}
\usepackage[bookmarks=false,colorlinks]{hyperref}
\hypersetup{
    linkcolor=magenta,        
    citecolor=PineGreen,     
    filecolor=Plum,         
    urlcolor=blue,           
}
\usepackage{graphicx}
\usepackage[protrusion=true,expansion=true,final]{microtype}



\begin{document}
\title{Strain-induced isosymmetric ferri-to-ferroelectric transition
with large piezoelectricity}
%

\author{Gaoyang Gou}\email{gougaoyang@mail.xjtu.edu.cn}
\affiliation{Frontier\!  Institute  of  Science  \& Technology, Xi'
an Jiaotong University, Xi' an  710049, China}
\author{James M.\ Rondinelli}
  \email{jrondinelli@coe.drexel.edu}
    \affiliation{Department\!  of\!  Materials\!  Science\!  \& Engineering,\!
    Drexel University,\! Philadelphia,\! PA 19104,\! USA}%
\date{\today}
\begin{abstract}
We identify a first-order, isosymmetric transition between a
ferrielectric (FiE) and ferroelectric (FE) state in $A$-site ordered
LaScO$_{3}$/BiScO$_{3}$ and LaInO$_{3}$/BiInO$_{3}$ superlattices.
Such a previously unreported ferroic transition is driven by the
easy switching of cation displacements without changing the overall
polarization direction or crystallographic symmetry.
Epitaxial strains less than 2\% are predicted to be sufficient to
transverse the phase boundary, across which we capture a $\sim\!5$X
increase in electric polarization. In a fashion similar to classic 
Pb-based perovskite ceramics with a morphotropic phase boundary
(MPB), we predict a large electromechanical response up to 131 pC/N 
in the vicinity of FiE-FE phase boundary.
We propose this transition as alternative ferroic
transition to obtain large piezoelectricity, with the additional
advantage of operating under extreme conditions in benign 
chemistries and without chemical disorder.
\end{abstract}
%
\maketitle \sloppy


The best piezoelectric materials belong to the functional
$AB$O$_{3}$ perovskite oxide ceramic class, where a small external
stimuli (mechanical stress or electric field) triggers a ferroic
transition with large changes to the overall polarization.
For the past few decades, the conventional approach to obtain high
piezoelectricity has relied on synthesizing ferroelectric (FE) solid
solutions in close proximity to a \emph{composition-induced}
morphotropic phase boundary (MPB). \cite{Lines:book} A large
piezoelectric response occurs at those boundaries, across which a
\emph{rotation} of polarization occurs between symmetry unique polar
phases.\cite{Guo/Shirane:2000,Fu/Cohen:2000}
Most MPB engineering requires alloying of different FE
end-members, which introduces complexity in deciphering the
microscopic origins necessary for property
optimization.\cite{ahart:2008,zeches:2009}
Therefore, identification of alternative ferroic transitions in a
\emph{single-phase} material suitable for generating large
piezoelectricity based on new mechanisms are critical to integrating
novel electroceramics into applications ranging from ultrasound
imaging technology to precision actuation in micromotors.
\cite{Uchino:book}
The recent discovery of a class of hybrid improper ferroelectrics
(HIF),\cite{Bousquet08,Benedek/Fennie:2011} where an 
electric polarization is driven by two coupled octahedral-rotation
lattice modes, offer opportunities for design of ferroic materials
with diverse functionalities.
Typically, in $A$-site ordered ultra-short period 1/1
ABO$_3$/A$^\prime$BO$_3$ perovskite superlattices exhibiting HIF,
the electric dipoles of each cation sublattices are restricted by
the symmetry of the orthorhombic $a^-a^-c^+$ $B$O$_6$ rotation
patterns \cite{Rondinelli/Fennie:2012} in Glazer notation
\cite{Glazer:1972}.
Such a rotocrystalline anisotropy stabilizes a \emph{ferrielectric}
(FiE) state,
\cite{Cross:1956,Suzuki/Okada:1978,Tagantsev:1987,Shimakawa:1999}
where two of the three cation-sublattice dipoles are directed along
the pseudo-cubic [$110$], while the third is anti-parallel to the
others [$\bar{1}\bar{1}0$], yielding a net non-zero electric
polarization. \cite{Benedek:2012}
In such a layered structure, the electric polarizations of each
cation-sublattice can be tuned independently---the $A$ and $B$
cation polarizations directly rely on the octahedral
connectivity of the system. \cite{Lopez-Perez/Iniquez:2011} 
\textit{Is there
a suitable way to tune the oxygen lattice topology and the
polarization direction of each cation-sublattice?}
If answered in the affirmative, then the FiE state   can be
directly converted into a FE state where all cation-sublattice
dipoles are in parallel alignment. Consequently, a new FiE-to-FE
ferroic transition would achieve a significant polarization enhancement
\emph{and} large piezoelectric response in \emph{cation ordered}
oxides.
In this paper, we use first-principles density functional
calculations to demonstrate that the \emph{polarization switching}
between a FiE and FE phases in HIF ABO$_3$/A$^\prime$BO$_3$
perovskite superlattices is accessible with substrate-induced epitaxial strain.
Specifically, we uncover a first-order isosymmetric phase transition
and evaluate the feasibility of the FiE-to-FE transition through a
minimum energy path (MEP) trajectory.
Similar to Pb-based solid-solutions, where a large piezoelectric
response occurs near the MPB driven by rotation of polarization along
an easy \emph{rotation path} between the rhombohedral and tetragonal
FE phases,\cite{Noheda_etal:2001,Grinberg02p909} 
we predict a large piezoelectric response---up to 131 pC/N 
around the FiE-FE
phase boundary, as a result of the easy \emph{polarization
switching} between FiE and FE phases.
Lastly, 
the FiE-to-FE transition has the practical
advantage of surviving to high temperature and at integral
stoichiometry in an ordered superlattice, without introducing
compositional complexity, toxicity, or a change in symmetry---features absent
in conventional Pb-based piezoelectrics.
\section*{Results}

\textbf{Models}.---%
We select $A$-site ordered ultra-short period 1/1
La$M$O$_3$/Bi$M$O$_3$ perovskite superlattices, with trivalent
cations $M=\textrm{Ga, Sc}$ and In, respectively, to guarantee the
insulating behavior required to sustain an electric polarization.
%
%
The $6s^2$ stereochemical lone pair of Bi
promotes structural distortions and enhanced polarizability
\cite{Seshadri/Hill:2001}.
Lastly, each bulk $A^{3+}M^{3+}$O$_3$ oxide forming the
superlattices exhibits the targeted orthorhombic $a^{-}a^{-}c^{+}$
octahedral rotation pattern [Fig.~S1 of the Supplementary Information (SI)] reported to
promote HIF \cite{Rondinelli/Fennie:2012}. 

\begin{figure}
\includegraphics[width=0.45\textwidth]{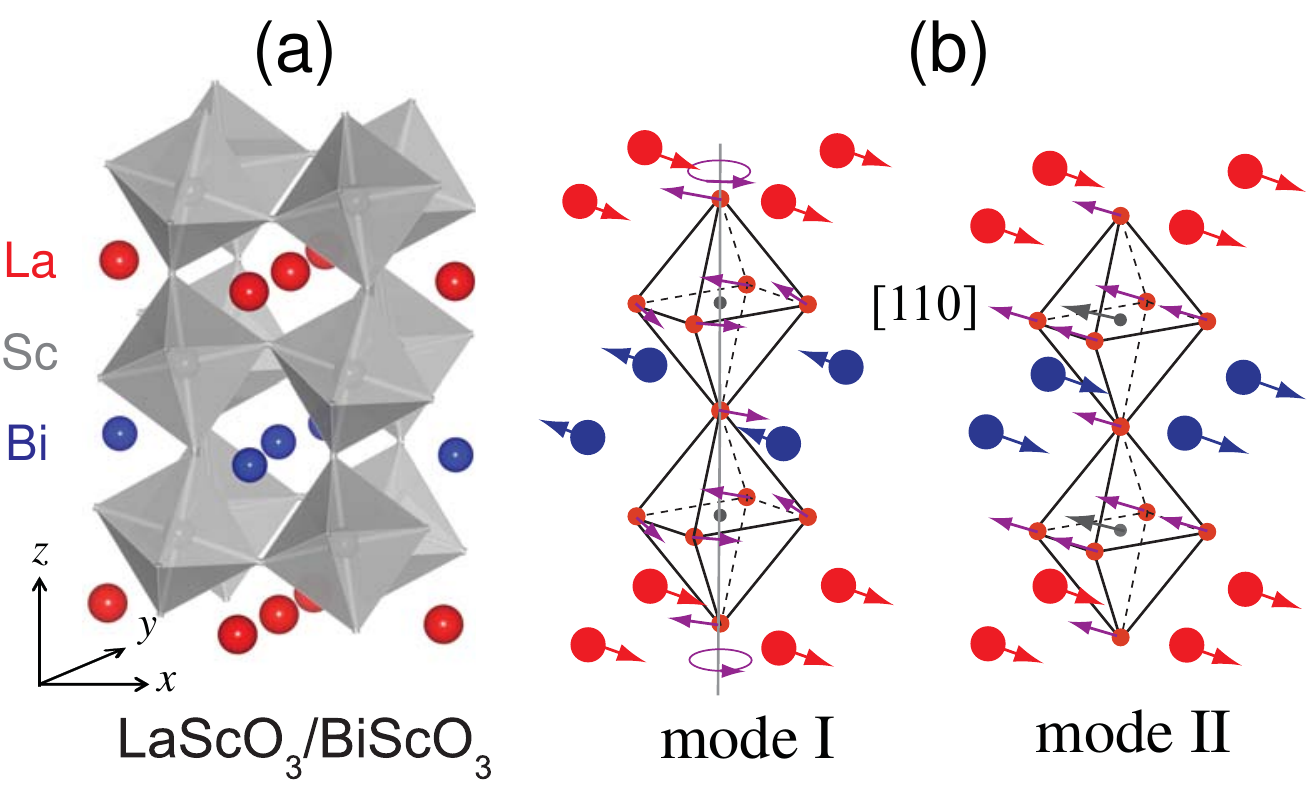}\vspace{-8pt}
\caption{(a) The $Pmma$ paraelectric structure of the
LaScO$_{3}$/BiScO$_{3}$ superlattice with the $a^{-}a^{-}c^{0}$
rotation pattern (a).
The displacement patterns of the corresponding
soft phonon modes (b):
$\omega_\textrm{I} $ with \emph{anti-parallel} La-O/Bi-O
in-plane displacements (left), and
$\omega_\textrm{II}$ with \emph{parallel} La-O/Bi-O in-plane displacements (right).} \label{Fig.1}
\end{figure}

\textbf{Ground State Structures}.---%
We search for the ground state structures by computing the phonons
in the Brillouin zone of the parent $P4/mmm$ phase, \emph{i.e.}, the
distortion-free $A/A^\prime$ cation ordered superlattice.
We find three principal unstable vibrational phonons in each superlattice:
%
They are characterized as either in-phase (irreducible
representation $M_3^+$ relative to $Pm\bar{3}m$) or out-of-phase
rotations ($R_4^+$) of O$_6$ octahedra, or polar inversion
symmetry lifting cation displacements along [110] ($\Gamma_4^-$),  
respectively (Fig.~S2 and Table SII).
For all superlattices, the most unstable modes are out-of-phase
rotations, $a^-a^-c^0$, with the mode-frequency of the gallate
smaller than the scandate or indiumate owing to the larger tolerance
factor of the former \cite{Mulder/Rondinelli/Fennie:2013}. 

We next freeze-in each soft mode into the $P4/mmm$ superlattices
followed by full structural relaxations to obtain lower symmetry
distorted phases.
For each superlattice, we find a stable non-polar $Pmma$ phase
with the $a^-a^-c^0$ rotation [\autoref{Fig.1}(a)].
We now take this as the relevant paraelectric phase for which
transitions between lower-symmetry polar phases could occur.
We find two unstable vibrational modes in the $Pmma$ superlattices
(Table SIII): $\omega_\textrm{I}$ and
$\omega_\textrm{II}$ correspond to atomic displacement vectors that
are either a combination of in-phase octahedral rotations and
anti-parallel $A$ cation displacements, or cooperative polar cation
displacements moving in the direction opposite of the oxygen anions
[\autoref{Fig.1}(b)], respectively. 

\begingroup
\squeezetable
\begin{table}[]
\begin{ruledtabular}
\caption{\label{Table1}%
The relative energetic stability (in meV/f.u.) with respect to the
ground state [$Pmc2_1$]$_\textrm{I}$ structure for various
low-energy phases specified by Glazer tilt; subscripts indicate the
direction of the electric polarization following the convention in
Ref~\onlinecite{Stokes/Kisi_et_al:2002}.
%
The
average tolerance factor ($\bar{\tau}$) for each perovskite
superlattice is given in parentheses.}
\begin{tabular}{lcccc}
                           & $Pmma$  & [$Pmc2_1$]$_\textrm{I}$                 & [$Pmc2_1$]$_\textrm{II}$                &  $Pc$                           \\
\cline{2-5}\\[-1.2em]
System / Rotation                           & $a_{0}^{-}a_{0}^{-}c_{0}^{0}$ & $a_{+}^{-}a_{+}^{-}c_{0}^{+}$  & $a_{+}^{-}a_{+}^{-}c_{0}^{+}$  &  $a_{+}^{-}a_{+}^{-}c_{+}^{-}$     \\[0.2em]
\hline
LaInO$_{3}$/BiInO$_{3}$ (0.86)     & 437                  & 0      &109             & 179                  \\
LaScO$_{3}$/BiScO$_{3}$ (0.89)     & 338                  & 0      &57              & 170                  \\
LaGaO$_{3}$/BiGaO$_{3}$ (0.94)     & 97                   & 0      & $-$            & $-$                  \\
\end{tabular}
\end{ruledtabular}
\end{table}
\endgroup

We now combine the $Pmma$ structures with the displacement vectors
for $\omega_\textrm{I}$ and $\omega_\textrm{II}$.
The symmetries of the superlattices reduce to the \emph{same}
polar space group $Pmc2_1$ (no.\ 26), where the polar axis
is along the pseudocubic [110] direction. 
%
%
Hereafter, we distinguish between the two symmetry equivalent $Pmc2_1$ phases
as (I) or (II), appended as a subscript to the space group label.
After performing a full structural relaxation, we find that the
[$Pmc2_1$]$_\textrm{I}$ phase is the global ground state for all
superlattices (\autoref{Table1}).
Except for LaGaO$_{3}$/BiGaO$_{3}$,  [$Pmc2_1$]$_\textrm{II}$ is
the next lowest energy phase, \emph{e.g.}, 57~meV per formula unit (f.u.) higher
in energy than [$Pmc2_1$]$_\textrm{I}$ in LaScO$_{3}$/BiScO$_{3}$.
\footnote{LaScO$_{3}$/BiScO$_{3}$ and LaInO$_{3}$/BiInO$_{3}$ also
possess a low energy polar $Pc$ structure with a polarization component
along [001]; however, the structures are more than 170~meV/f.u.\ higher
energy and likely inaccessible via substrate-induced epitaxial
strain at room-temperature. Therefore, we do not consider 
it further.} 
The small energy difference suggests transition between the two
$Pmc2_{1}$ phases may be achieved by thin film epitaxy.

\begin{figure}[b]
\centering
\includegraphics[width=0.92\columnwidth]{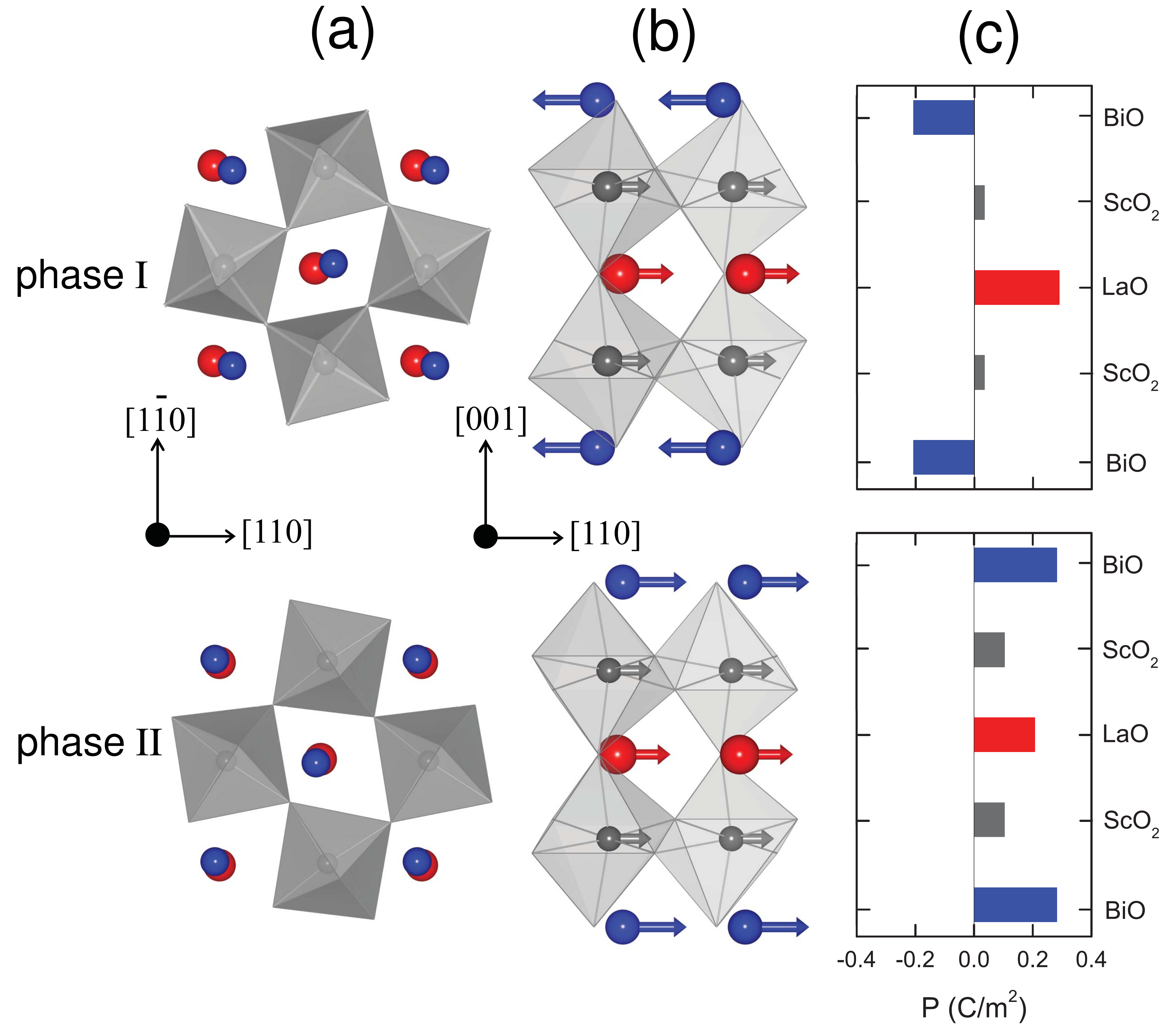}\vspace{-8pt}
\caption{Atomic structures of the two isosymmetric polar ground
states ($a_{+}^{-}a_{+}^{-}c_{0}^{+}$) with $Pmc2_{1}$ symmetry:
phase I (upper) and phase II (lower).
%
Both top view (a) and side view
(b) are shown to emphasize cation displacement directions (arrows).
(c) Layer-revolved polarization
$\vec{p}_\mathrm{layer}=\Omega^{-1}\sum_{i}{u_{i}Z^{*}_{i}}$, where $\Omega$
is the cell volume, $u_{i}$ is the $i$ cation displacement
and $Z^{*}_{i}$ is the Born effective charge.
Phase I  is \emph{ferrielectric}
(total polarization, $P =0.12$~C/m$^{2}$) with anti-parallel $\vec{p}_\textrm{LaO}$ and
$\vec{p}_\textrm{BiO}$, while Phase II is a normal \emph{ferroelectric}
($P = 0.63$~C/m$^{2}$), where both $\vec{p}_\textrm{LaO}$ and
$\vec{p}_\textrm{BiO}$ are oriented in the same direction.}\label{Fig.2}
\end{figure}

\begin{figure}
\centering
\includegraphics[width=0.85\columnwidth]{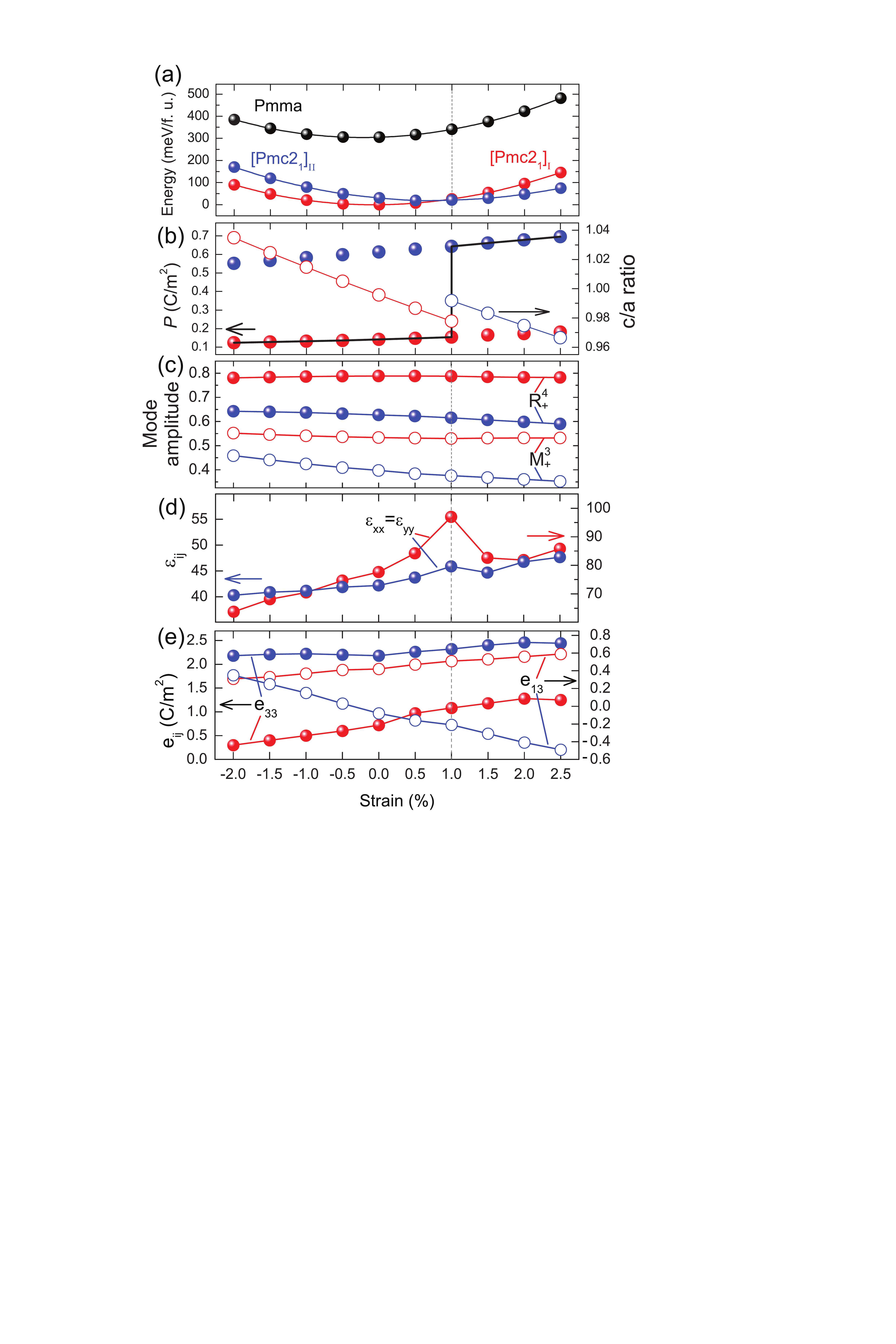}\vspace{-8pt}
\caption{Strain-dependent  phase stability and electronic properties
of LaScO$_{3}$/BiScO$_{3}$: (a) total energies for the paraelectric
and polar  structures, (b) total polarization [in filled circles,
the bold black line transverses the polarization of the lowest
energy structures in (a)] and $c/a$ axial ratio (in open circles), (c)
the out-of-phase tilting ($R^{+}_{4}$ in filled circles) and
in-phase rotation ($M^{+}_{3}$ in empty circles) amplitudes, (d)
static dielectric tensor $\epsilon_{ij}$
($\epsilon_{xx}$=$\epsilon_{yy}$, filled circles), and (e) the
piezoelectric stress constants $e_{ij}$ ($e_{33}$, filled circles;
$e_{31}$ empty circles) for $Pmc2_{1}$ phase I (red) and II (blue).
The isosymmetric transition occurs at 1.0 \% tensile strain (broken
lines).} \label{Fig.3}
\end{figure}

To understand the energy difference between the two isosymmetric
$Pmc2_1$ phases, we focus on the LaScO$_{3}$/BiScO$_{3}$
superlattice, and examine its local atomic structures in more
detail. \autoref{Fig.2}(a) and (b) shows both polar phases projected
along the [100] and [011] directions.
Each possesses the same $a^{-}a^{-}c^{+}$
rotations, but \emph{different cation displacement patterns},
which become readily apparent in the 
layer-resolved electric polarizations [\autoref{Fig.2}(c)].
[$Pmc2_1$]$_\textrm{I}$ is described as a \emph{ferrielectric}
state, whereby a non-cancellation of anti-aligned layer dipoles
($\vec{p}_\mathrm{layer}$) from the alternating LaO or BiO sublattices in
the superlattice produces an overall non-zero electric polarization.
In [$Pmc2_1$]$_\textrm{II}$, $\vec{p}_\textrm{BiO}$ switches
from anti-parallel to parallel alignment with $\vec{p}_\textrm{LaO}$.
All cations displace in the same direction, as in a conventional
\emph{ferroelectric}, to give a more than 5X polarization enhancement. 

[$Pmc2_1$]$_\textrm{I}$ is obtained by condensation of the
$\omega_{I}$ soft mode, and  
the polar displacements are stabilized from a combination of the
$a^{-}a^{-}c^{0}$ and $a^{0}a^{0}c^{+}$ octahedral rotations
(Fig.~S1).
The polarization in [$Pmc2_1$]$_\textrm{II}$ arises from the polar
instability $\omega_\textrm{II}$ in $Pmma$, where cooperative displacement
of \emph{all} cations occur in the same direction and are opposite
to the oxygen anions, as found in proper ferroelectrics,
\emph{e.g.}, BaTiO$_{3}$.
Moreover, phase II has octahedral rotation amplitudes that are
\emph{suppressed} compared to phase I [\autoref{Fig.3}(c)] in the
presence of a greater electric polarization, indicating the
trilinear coupling term connecting the octahedral rotations to the
polarization contributes less to the ground state
\cite{Benedek/Fennie:2011,Mulder/Rondinelli/Fennie:2013}.
Unlike the low-temperature  FiE-to-FE  transitions
observed in smectic liquid-crystalline systems, 
\cite{nishiyama:2001,hiraoka:1990,yamashita:1996} 
where  chirality or azimuthal reorientation of  molecular units 
occurs via weak intermolecular forces,\cite{Lorman:1994}  
in these oxide superlattices an extraordinary large energy gain
derives from the steric stability of the octahedral rotations.
For example, 338~eV/f.u.\ between the polar and paraelectric states of LaScO$_{3}$/BiScO$_{3}$ 
\emph{viz.}\ 48~meV/f.u.\ for BaTiO$_{3}$,\cite{Sanna/Schmidt:2011} which 
has a Curie temperature ($T_c$) of $\sim$120$^\circ$.
As a consequence, these perovskite superlattices should have much 
higher ferroelectric $T_c$'s ($>$500$^\circ$C) and  
persistent electric polarizations, similar to that reported for other HIF oxides
\cite{Lopez-Perez/Iniquez:2011}.


\textbf{Strain-induced FiE-to-FE Transition}.---%
The evolution of the $Pmc2_1$ phases with applied epitaxial strain
are examined next. \autoref{Fig.3}(a) displays our calculated total
energy for the two phases with respect to strain. We find in
LaScO$_{3}$/BiScO$_{3}$, the transition from [$Pmc2_1$]$_\textrm{I}$
to [$Pmc2_1$]$_\textrm{II}$ occurs approximately at 1.0\% tensile
strain. Such a transition is first-order and isosymmetric;
\cite{Rondinelli/Coh:2011} it is accompanied by a strong
discontinuous increase in the overall polarization and $c/a$ axial
ratio and a reduction of the octahedral rotation amplitudes
[\autoref{Fig.3}(b) and (c)].
The FiE-to-FE transition also occurs in the indium superlattice,
albeit at a larger strain ($\simeq$2.0\%), yet it is not found in
the gallium compound (Fig.~S5).
The larger tolerance factor for the
LaGaO$_{3}$/BiGaO$_{3}$ reduces the energy gain between
$Pmma$ and [$Pmc2_1$]$_\textrm{I}$  (\autoref{Table1}),
and makes the [$Pmc2_1$]$_\textrm{II}$ phase unstable.
In the remainder of this paper, we focus on the strained
LaScO$_{3}$/BiScO$_{3}$ with an emphasis on property response 
enhancement near the FiE-to-FE transition.

\begingroup
\squeezetable
\begin{table}[b]
\begin{ruledtabular}
\caption{\label{Table2}%
Calculated piezoelectric stress constants $e_{ij}$ (C/m$^{2}$)
and piezoelectric strain coefficients $d_{ij}$ (pC/N) in 
LaScO$_{3}$/BiScO$_{3}$ at 1.0\% tensile strain for
single (mono-domain) 
[$Pmc2_{1}$]$_\textrm{I}$ (FiE) 
and [$Pmc2_{1}$]$_\textrm{II}$ (FE)
phases, and a multi-domain  FiE-FE structure.}
\begin{tabular}{lcccccc}
   & \multicolumn{2}{c}{[$Pmc2_{1}$]$_\textrm{I}$} & \multicolumn{2}{c}{[$Pmc2_{1}$]$_\textrm{II}$}  & \multicolumn{2}{c}{multi-domain}  \\
   \cline{2-3}\cline{4-5}\cline{6-7}
index ($ij$)      &  $e_{ij}$  &  $d_{ij}$   &  $e_{ij}$  &   $d_{ij}$  &  $e_{ij}$  &   $d_{ij}$  \\
\hline
31          &   0.50     &  -1.05      &  -0.19     &  -6.16    & -4.82   &  -59.72   \\
32          &   1.05     &   2.33      &   0.77     &  0.91     &  5.01   &   18.53   \\
33          &   1.34     &   5.23      &   2.45     &  15.66    &  8.91   &   96.93   \\
24          &   0.93     &   14.57     &   1.03     &  17.96    &  7.13   &  110.76   \\
15          &   1.10     &   14.30     &   1.23     &  21.61    &  5.03   &  131.20   \\
\end{tabular}
\end{ruledtabular}
\end{table}
\endgroup

The evolution of the response functions, including the dielectric
permittivity and piezoelectric response, as a function of strain are
computed at the DFT-level \cite{Wu/Vanderbilt:2005}.
\autoref{Fig.3}(d) shows that the in-plane dielectric susceptibility
$\varepsilon_{xx}$ and $\varepsilon_{yy}$ are maximum at the
FiE-to-FE transition at 1.0 \% tensile strain in both $Pmc2_{1}$
phases, consistent with the first-order behavior \cite{Wu/Cohen:2005}.
%
%
In \autoref{Fig.3}(e), we plot the piezoelectric stress constants
$e_{33}$ and $e_{13}$, which evolve smoothly with respect to strain.
The non-zero piezoelectric strain coefficients $d_{ij}$ for the
\emph{individual} mono-domain FiE and FE phases at 1.0\% strain are also
reported in \autoref{Table2}.
With the presence of relative small polarization and large
octahedral rotation amplitudes, each polar phases displays a modest
piezoelectric response, compared to that of proper FE oxides without
octahedral rotations (\emph{e.g.}, PbTiO$_{3}$).
Similar piezoelectric responses have been reported in
other hybrid-improper ferroelectric oxides as
well.\cite{Lopez-Perez/Iniquez:2011}
%
Despite the small piezoelectric moduli found in the
\emph{individual} FiE and FE phases, one should also consider the
possible enhancements due to the FiE-to-FE \emph{transition} in describing
the macroscopic functionality.

\begin{figure}[b]
\hspace*{-4pt}\includegraphics[width=0.485\textwidth]{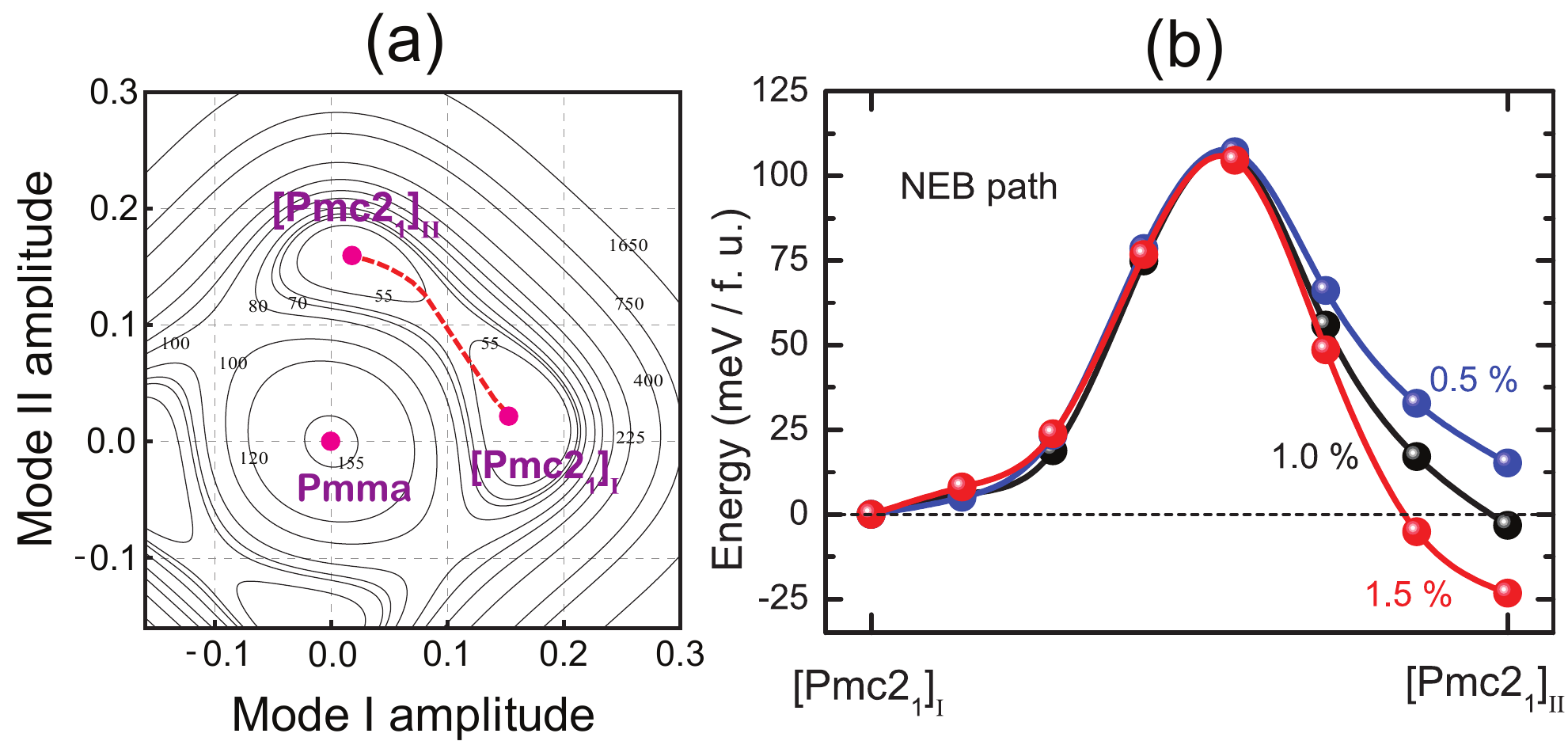}\vspace{-5pt}
\caption{
The calculated two-dimensional energy surface contours in meV for
LaScO$_{3}$/BiScO$_{3}$ with respect to the soft phonons of
$Pmma$ (a).
The MEP between
two polar states for LaScO$_{3}$/BiScO$_{3}$ under 
tensile strain (b). 
\label{Fig.4}}
\end{figure}

\textbf{Piezoelectricity at the FiE-to-FE Transition}.---%
To 
evaluate the consequence of the FiE-to-FE transition on
piezoelectricity, we first map out the energy landscape in the
vicinity of the low energy phases (\autoref{Fig.4}).
The two-dimensional energy contours at 1.0\% strain for
LaScO$_{3}$/BiScO$_{3}$ are computed by varying the amplitude of the
eigenvectors of the unstable $Pmma$  phonons.
We find that the FiE [$Pmc2_1$]$_\textrm{I}$ and FE
[$Pmc2_1$]$_\textrm{II}$ phases are located at 
minima near the major axes, separated by a energy barrier.
We estimate the feasibility of the FiE-to-FE transition by
calculating that barrier separating the two stable structures
along a MEP obtained from the NEB method
[\autoref{Fig.4}(b)].
We find a small, nearly strain-independent 106 meV/f.u.\ barrier separating the phases,
\footnote{The height of
the energy barrier is also consistent with that obtained from the
energy contour plot in \autoref{Fig.4}.}
and 
comparable to 
the polarization rotation barrier in piezoelectric PbTiO$_3$
\cite{Hong/Vanderbilt:2011}, suggesting the FiE-to-FE transition
should be energetically accessible. \footnote{The experimental
switching path will depend on many complicated extrinsic factors,
including the coexistence of two mono-domains, motion of the
domain-wall. However, our calculated energy barrier of the
mono-domain crystal provides a reasonable estimate for the most
likely intrinsic polarization switching scenario in the strained
LaScO$_{3}$/BiScO$_{3}$ superlattice.}

\begin{figure}[b]
\centering
\includegraphics[width=0.87\columnwidth]{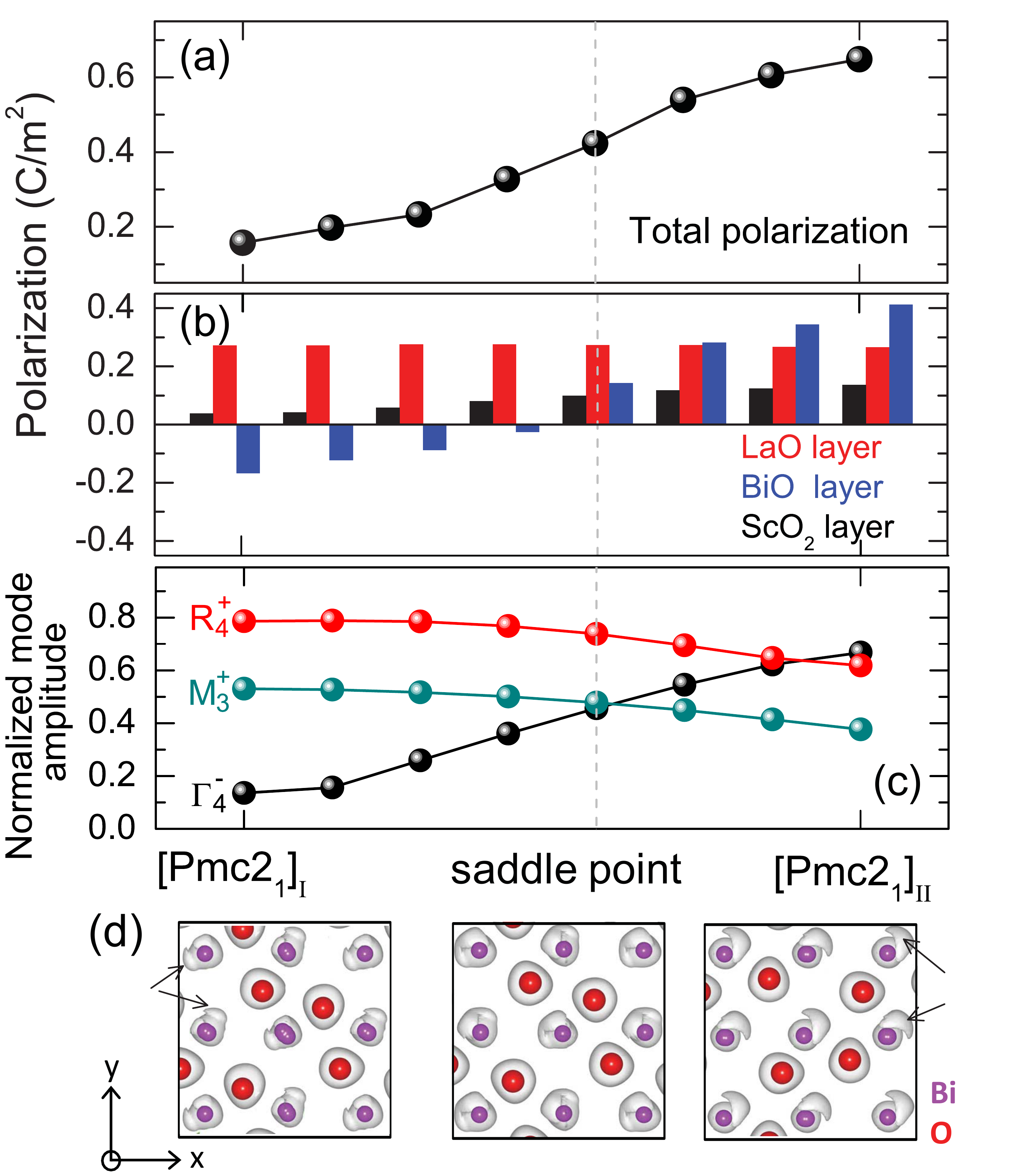}\vspace{-3pt}
\caption{The calculated (a) total polarization, (b) layer-revolved
polarizations and (c) major crystallographic mode amplitudes for
each image along the MEP connecting phase I and II in
LaScO$_{3}$/BiScO$_{3}$ at 1.0\% strain.
The position of the saddle point along the MEP is shown as dashed line.
Electronic-localization-function (ELF) surfaces within the BiO
layer for phase I, the intermediate structure at the saddle point,
and phase II (d).
Across the FiE-to-FE transition, the Bi lone pair  
alignment reverses (small lobe, arrowed) .
} \label{Fig.5}
\end{figure}

We emphasize that the change of polarization across the
$[Pmc2_1]_\textrm{I}\rightarrow [Pmc2_1]_\textrm{II}$ transition
[\autoref{Fig.5}(a)] along the MEP is mainly due to a rigid
\emph{switching} in direction of Bi displacements along [110] direction
above and below the ScO$_{2}$ layers (\autoref{Fig.2}).
Across the phase-transition, $\vec{p}_\textrm{BiO}$ switches from
anti-parallel to parallel alignment with $\vec{p}_\textrm{LaO}$
along the polar axis by decreasing its amplitude in the (001)-plane
until reaching the saddle point in the energy landscape
[\autoref{Fig.5}(b)]. Here, a nearly random distribution of the Bi-lone 
pairs within the planes emerges [\autoref{Fig.5}(d)].
We find a transformation of BiO$_{8}$ polyhedra into highly
distorted BiO$_{4}$ tetrahedra with enhanced Bi$^{3+}$ electron lone
pair activity in $[Pmc2_1]_\textrm{II}$ (Fig.~S3 and Fig.~S4).
After the saddle point, $\vec{p}_\textrm{BiO}$ continues to increase
while $\vec{p}_\textrm{LaO}$ remains nearly unchanged.
Critically, no octahedral rotation-reversal is necessary across this
unusual ferroic transition; only a moderate reduction of
$\simeq$25\% in the rotation angle amplitude is required
[\autoref{Fig.5}(c)].
%
%
Specifically the MEP is dominated by a large enhancement of
$\Gamma_{4}^{-}$ polar displacements ($>$5X), while the
\emph{polarization direction remains unchanged}.
This behavior is unique from the polarization-rotation phenomenon
commonly observed in FE oxide solid solutions with MPB,
\cite{Fu/Cohen:2000} where the polarization vector aligns along
different crystallographic directions.
The structure evolution along the polarization-switching path
indicates the FiE state is directly converted into FE state without
passing through any paraelectric intermediate and large polarization
enhancement can be achieved through the phase-transition. Therefore
a large piezoelectric response is expected in the vicinity of the FiE-FE 
boundary \cite{Fu/Cohen:2000}.

Since the two phases are connected by a first-order phase
transition, it is experimentally likely that both the FiE and FE
phases would coexist and domain-walls separating them will be
formed in tensile-strained LaScO$_{3}$/BiScO$_{3}$ films,
\emph{e.g.}, those grown epitaxially on a (001)-terminated LaLuO$_3$
substrate with a pseudo-cubic lattice constant of 4.17~{\AA}.
Thus to realistically predict the piezoelectric response across the FiE-to-FE transition, the 
simulation of a \emph{multi-domain} structure and various domain 
wall (DW) configurations are necessary.
We have explored several low-Miller indice planes with 
neutral and mechanically compatible DWs to limit the number of 
feasible configurations to only those that are energetically 
most likely to occur (Fig.\ S6 and Table SVII).
Owing to the enormous computational overhead, 
we describe next only the change in layer polarizations and enhanced piezoelectric 
coefficients for  the lowest energy multi-domain structure.

\begin{figure}
\centering
\includegraphics[width=1\columnwidth]{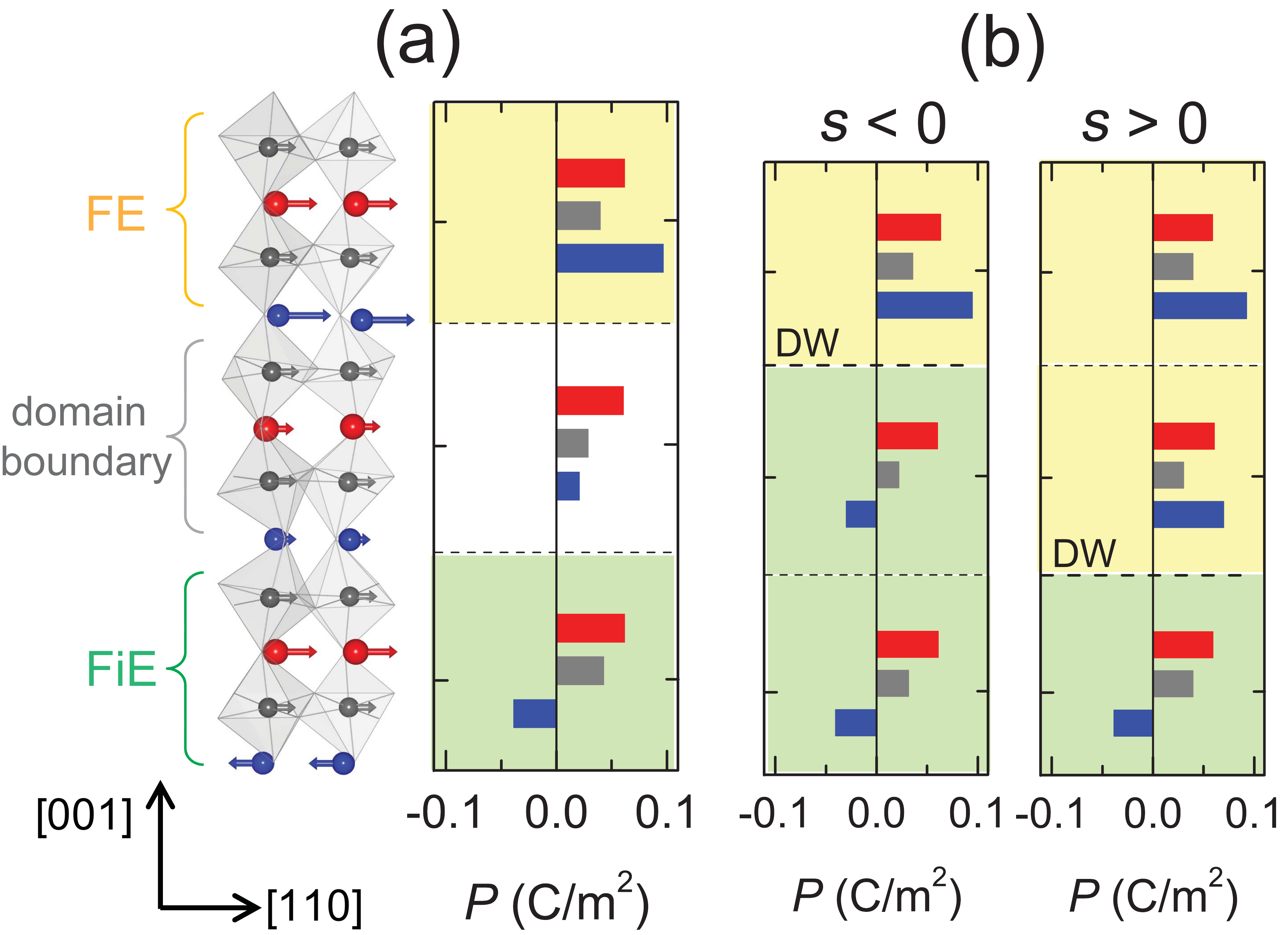}\vspace{-7pt}
\caption{Equilibrium atomic structure for a multiple domain  LaScO$_{3}$/BiScO$_{3}$ 
superlattice containing FiE and FE single domains separated by a 0$^\circ$-(001) 
domain wall (DW). 
(a) The distribution of cation layer-revolved polarizations 
($\vec{p}_\textrm{BiO}$ in blue, 
 $\vec{p}_\textrm{LaO}$ in red, and 
 $\vec{p}_\textrm{ScO$_2$}$ in gray) across
the multi-domain structure at 0\% strain. 
The FiE and FE mono-domains states with 
$\vec{p}_\textrm{BiO}$ in either an anti-parallel or parallel alignment with 
the global [110]  polarization direction  
are  shaded in green and yellow colors, respectively.
There is the abrupt change of $\vec{p}_\textrm{BiO}$ (blue bar) 
at the boundary (broken line) between the two domains 
(unshaded region). 
(b) The motion of the domain wall 
and layer polarization distribution for the multi-domain configuration under 
1.0\% compressive ($s<0$) and tensile ($s>0$) strain
applied along the [110] direction.} \label{Fig.6}
\end{figure}

\autoref{Fig.6}(a) shows the relaxed FiE-FE
multi-domain configuration for the lowest energy  0$^\circ$-(001) 
domain wall structure.
It consists of two distinct FiE and FE domains (shaded regions), 
which have atomic structures and layer polarizations 
consistent with the mono-domain FiE and FE phases 
described earlier. 
No reorientation of the overall
electric polarization direction or octahedral rotation 
sense occurs in this multi-domain state, indicating  that 
the simulation cell 
is sufficiently large to capture the multi-domain ferroelectric 
and piezoelectric responses.

The evolution of the layer-revolved polarizations indicates
that the DW structure contributes to the overall 
electric polarization by favoring \emph{parallel} alignment 
of all $A$O and $B$O$_2$ dipoles along the  [110] global-polarization direction 
in the boundary region.
The $\vec{p}_\textrm{BiO}$ direction \emph{switches} to align along  
this direction in the DW region, and remarkably, 
the displacement patterns in the boundary region are 
almost identical to that found at the saddle point along the MEP 
[\autoref{Fig.5}(b)]. 
Therefore, the boundary in a multi-domain FiE-FE microstructure may act 
as a two-dimensional ``piezoelectrically active'' region under an 
applied electric or mechanical field.

Our simulated piezoelectric coefficients for the multi-domain
structure are reported in \autoref{Table2}. 
Compared to the mono-domain FiE or FE
structures, both $e_{ij}$ and $d_{ij}$ tensors are significantly 
enhanced in the multi-domain system ($>$6X). 
The largest piezoelectric strain coefficient $d_{15}$ reaches 
131 pC/N, which is comparable or even larger than that of most 
single crystal perovskite oxides or digitally ordered superlattices 
exhibiting proper ferroelectricity. 
%
\footnote{Relevant values for comparison include:
$d_{33}\simeq70$~pC/N for PbTiO$_{3}$ \cite{PhysRevB.81.014102}; 
36~pC/N for BaTiO$_{3}$ \cite{PhysRevB.85.054108}; 
40~pC/N for KNbO$_{3}$ \cite{wan:104107}; 
54~pC/N for CaTiO$_{3}$/BaTiO$_{3}$ superlattices\cite{PhysRevLett.104.207601}; 
and 
100~pC/N for highly strained BiFeO$_{3}$ thin films\cite{strained_BFO}.} 
The large piezoelectric response found in the multi-domain configuration 
originates from the DW motion\cite{zhang:2012} under 
an applied mechanical strain  [\autoref{Fig.6}(b)].
We find that under compressive strain ($s<0$), 
$\vec{p}_\textrm{BiO}$ in the domain boundary region \emph{switches} 
from alignment along the [110] to [$\bar{1}\bar{1}0$] 
direction, whereas $\vec{p}_\textrm{layer}$ from the other $A$O and 
ScO$_2$ planes are weakly modified.
This FiE domain expands and dominates, creating a state 
with a layer polarization gradient.
The DW motion also occurs for applied tensile strains ($s>0$), except 
the FE domain grows at the expense of the FiE domain [greater yellow 
shaded region in \autoref{Fig.6}(b)].
We note that during the domain wall motion process, 
the cations always displace along the \emph{easy
polarization-switching} path as determined from the MEP between FiE and
FE phases [\autoref{Fig.5}].
%
Therefore, the isosymmetric FiE-to-FE transition with the easy
polarization-switching path in the $A$-site ordered oxide
superlattices manifests as an efficient, previously underutilized, 
ferroic transition to obtain large piezoelectricity in material systems beyond 
Pb-based piezoelectric ceramics.

\section*{Summary and Conclusions}

We use first-principles calculations combined with group theoretical
studies to uncover a first-order, isosymmetric, structural
transition between a FiE and FE state in artificial $A$-site ordered
perovskite superlattices.
For LaScO$_{3}$/BiScO$_{3}$, we predict 1\% tensile epitaxial strain
is sufficient to induce a FiE-to-FE transition with electrical
features that could enable rational technology design:
A perovskite superlattice with two polar states, processing a large
polarization-magnitude difference, and simultaneously large
piezoelectric responses driven by easy switching of cation
displacements without changes to the overall polarization direction
or crystallographic symmetry---features absent in conventional
ferroelectric electroceramics with MPB.
Second, owing to the extraordinary large energy gain between the polar and
paraelectric states, the FE-to-FiE transition should persist up to
high temperature ($T>500^\circ$C), where Pb-based piezoelectrics responses
are impaired.
Finally, ferrielectric materials should exhibit similar electrical
switching behavior as antiferroelectrics, \emph{e.g.}, double
hysteresis loops that improve energy density capacities
\cite{doi:10.1021/ma2024057}, yet with the advantage of operating at
much lower fields, enabling practical integration for energy
harvesting devices.
%

\section*{Methods}

\textbf{Theoretical calculations}.---%
Our density functional theory (DFT) calculations are performed with
the \textsc{quantum espresso} code  \cite{QuantumEsprsso:2009}, with
optimized nonlocal norm-conserving \cite{Rappe:1990,Ramer:1999}
pseudopotentials, a 60 Ry plane-wave cutoff.
The PBEsol exchange-correlation functional\cite{PBEsol:2008}
 is used in the calculations; it provides an improved description of the atomic structural results over
LDA or PBE (Table~SI).
The Brillouin zone (BZ) integrations use a 8$\times$8$\times$8
Monkhorst-Pack $k$-point grid \cite{Monkhorst/Pack:1976}.
%

Low-energy phases are obtained by performing complete structural
relaxations on soft-mode generated structures until the
Hellmann-Feynman forces on the atoms are less than 0.1
meV~\AA\!$^{-1}$. For the strained films, the out-of-plane lattice
constant is relaxed until the stresses are less than 0.1 kbar.
The FiE-FE multi-domain supercell is constructed by combining the 
mono-domain FiE and FE stuctures and separating them by adomain walls, 
creating a simulation cell  with a total of 60 atoms, \emph{i.e.}, 
six times of the pseudocubic unit cells.
Additional details of the domain wall configurations explored are found in 
the SI.
The electronic contribution to the polarization 
is calculated following the Berry phase formalism
\cite{King-Smith/Vanderbilt:1993}.
The minimum energy path (MEP) between the
 stable phases is simulated using the nudged elastic band (NEB)
method \cite{henkelman:9901,henkelman:9978}.

\begin{acknowledgments}
G.Y.G.\ was supported by the 
NBR Program of China (2012CB619401),  
NSFC (11204230) and acknowledges computational support from NSCC.
G.Y.G.\ thanks C. Wu and J. Shi for assistance in preparing this
manuscript.
J.M.R.\ was supported by ARO (W911NF-12-1-0133)
and thanks C.J.\ Fennie and A.T.\ Mulder for insightful discussions.
\end{acknowledgments}

\section*{Author contributions}
G.Y.G.\ performed the theoretical calculations. G.Y.G.\ and J.M.R.\
conceived of the research project, contributed to the analysis of
the data and writing of the article.

\section*{Additional information}
Supplementary information is available in the online version of the paper.
Reprints and permissions information is available online at
\url{www.nature.com/reprints}.
Correspondence and requests for materials should be addressed to
G.Y.G.

\section*{Competing financial interests}
The authors declare no competing financial interests.

\vspace*{5\baselineskip}


\end{document}